\documentclass[iop, twocolumn, twocolappendix]{aastex631}

\usepackage{lineno}
\usepackage{graphicx}

\usepackage{graphicx}
\usepackage{color}
\DeclareUnicodeCharacter{02BC}{'}

\received{XXX}
\revised{XXX}
\accepted{XXX}
\submitjournal{XXX}

\shortauthors{Radio pulse search of four magnetars and a magnetar-like pulsar}

\begin{document}
\title{Deep Searches for Radio Pulsations and Bursts from Four Magnetar and a Magnetar-like pulsar with FAST} 

\correspondingauthor{Na Wang}
\email{na.wang@xao.ac.cn}

\nocollaboration{1}

\author[0000-0002-1052-1120]{Juntao Bai}
\affiliation{Xinjiang Astronomical Observatory, Chinese Academy of Sciences, Urumqi, Xinjiang 830011, Peopleʼs Republic of China}
\affiliation{School of Astronomy and Space Science, University of Chinese Academy of Sciences, Beijing, 100049, Peopleʼs Republic of China}

\author[0000-0002-9786-8548]{Na Wang}
\affiliation{Xinjiang Astronomical Observatory, Chinese Academy of Sciences, Urumqi, Xinjiang 830011, Peopleʼs Republic of China}
\affiliation{Key Laboratory of Radio Astronomy, Chinese Academy of Sciences, Urumqi, Xinjiang 830011, Peopleʼs Republic of China}
\affiliation{Xinjiang Key Laboratory of Radio Astrophysics, 150 Science1-Street, Urumqi, Xinjiang 830011, Peopleʼs Republic of China}

\author[0000-0002-9618-2499]{Shi Dai}
\affiliation{CSIRO Space and Astronomy, PO Box 76, Epping, NSW 1710, Australia}
\affiliation{Western Sydney University, Locked Bag 1797, Penrith South DC, NSW 2751, Australia}

\author[0000-0003-4498-6070]{Shuangqiang Wang}
\affiliation{Xinjiang Astronomical Observatory, Chinese Academy of Sciences, Urumqi, Xinjiang 830011, Peopleʼs Republic of China}
\affiliation{Key Laboratory of Radio Astronomy, Chinese Academy of Sciences, Urumqi, Xinjiang 830011, Peopleʼs Republic of China}
\affiliation{Xinjiang Key Laboratory of Radio Astrophysics, 150 Science1-Street, Urumqi, Xinjiang 830011, Peopleʼs Republic of China}

\author[0000-0002-5381-6498]{Jianping Yuan}
\affiliation{Xinjiang Astronomical Observatory, Chinese Academy of Sciences, Urumqi, Xinjiang 830011, Peopleʼs Republic of China}
\affiliation{Key Laboratory of Radio Astronomy, Chinese Academy of Sciences, Urumqi, Xinjiang 830011, Peopleʼs Republic of China}
\affiliation{Xinjiang Key Laboratory of Radio Astrophysics, 150 Science1-Street, Urumqi, Xinjiang 830011, Peopleʼs Republic of China}

\author[0000-0002-7662-3875]{Wenming Yan}
\affiliation{Xinjiang Astronomical Observatory, Chinese Academy of Sciences, Urumqi, Xinjiang 830011, Peopleʼs Republic of China}
\affiliation{Key Laboratory of Radio Astronomy, Chinese Academy of Sciences, Urumqi, Xinjiang 830011, Peopleʼs Republic of China}
\affiliation{Xinjiang Key Laboratory of Radio Astrophysics, 150 Science1-Street, Urumqi, Xinjiang 830011, Peopleʼs Republic of China}

\author[0000-0002-9173-4573]{Lunhua Shang}
\affiliation{School of Physics and Electronic Science, Guizhou Normal University, Guiyang 550001, China}
\affiliation{Guizhou Provincial Key Laboratory of Radio Astronomy and Data Processing, Guizhou Normal University, Guiyang 550001, China.}

\author[0009-0006-3224-4319]{Xin Xu}
\affiliation{School of Physics and Electronic Science, Guizhou Normal University, Guiyang 550001, China}
\affiliation{Guizhou Provincial Key Laboratory of Radio Astronomy and Data Processing, Guizhou Normal University, Guiyang 550001, China.}

\author[0000-0002-2060-5539]{Shijun Dang}
\affiliation{School of Physics and Electronic Science, Guizhou Normal University, Guiyang 550001, China}
\affiliation{Guizhou Provincial Key Laboratory of Radio Astronomy and Data Processing, Guizhou Normal University, Guiyang 550001, China.}

\author{Zhen Zhang}
\affiliation{Xinjiang Astronomical Observatory, Chinese Academy of Sciences, Urumqi, Xinjiang 830011, Peopleʼs Republic of China}

\nocollaboration{11}

\begin{abstract}

We report on radio observations of four magnetars SGR 0501+4516, Swift 1834.9$-$0846, 1E 1841$-$045, SGR 1900+14 and a magnetar-like pulsar PSR J1846$-$0258 with the Five-hundred-meter Aperture Spherical radio Telescope (FAST) at 1250 MHz. Notably, PSR J1846$-$0258 was observed one month after its 2020 X-ray outburst. The data from these observations were searched for periodic emissions and single pulses. No radio emission was detected for any of our targets. After accounting for the effect of red noise, the non-detections yield stringent upper limits on the radio flux density, with $S_{1250} \leq 16.9\, \mu $Jy for the four magnetars and the magnetar-like pulsar, along with constraints on single-pulse flux densities. Our deep radio observations suggest that these magnetars and the magnetar-like pulsar are indeed radio-quiet sources or unfavorably beamed. The resulting flux upper limits, along with previous findings, are discussed, highlighting the significance of further radio observations of radio-quiet magnetars and the high-B magnetar-like pulsar.

\end{abstract}


\section{Introduction} \label{sec:intro}

Magnetars are believed to be neutron stars with extremely strong magnetic fields~\citep[e.g.,][]{kaspi2017ARA&A}. Despite their energetic and extreme activities in X-ray and $\gamma$-ray, so far only six magnetars have been observed to be radio-active~\citep[1E 1547.0$-$5408, PSR J1622$-$4950, SGR J1745$-$2900, XTE J1810$-$197, Swift J1818.0$-$1607 and SGR J1935+2154;][]{camilo2007ApJ, levin2010ApJ, shannon2013MNRAS, halpern2005ApJ, karuppusamy2020ATel, scholz2020ATel, Wang2023arXiv230808832W, Zhu2023SciA}. 
The origin of radio emissions from magnetars is still uncertain and potentially carries key information on the nature of these unique objects. 
The radio emission of a magnetar may not be solely driven by its strong magnetic field but could also be influenced by the star's rotation, similar to the mechanisms observed in ordinary radio pulsars \citep{rea2012ApJ}.  
\citet{esposito2020ApJ} shows that the balance between magnetic energy and rotational power might differ considerably between different sources of the same class. The rotational energy loss rate of radio magnetars significantly exceeds their quiescent X-ray luminosity \citep{rea2012ApJ, Olausen2014ApJS}.  Some theories draw parallels between magnetar radio emissions and those of ordinary pulsars.  Pulsars emit radio waves through mechanisms such as polar cap acceleration and outer gap acceleration~\citep{Szary2015ApJ}. Another theory posits that magnetic reconnection events in the magnetosphere could be responsible for the radio emissions \citep{Lyubarsky2020ApJ}.

Closely linked with magnetars is a small population of rotation-powered, young pulsars with high surface magnetic fields ($\sim10^{13}$\,G), known as high-B pulsars. 
Magnetar-like X-ray activities have been detected from high-B pulsars J1846$-$0258 and J1119$-$6127 \citep{Gavriil2008, archibald2016ApJ}, which suggests that strong magnetic fields are indeed powering X-ray bursts. Interestingly, while J1119$-$6127 is an active radio pulsar~\cite{wjm+10,wje+11} that often exhibits enhanced and strongly variable radio emission following X-ray outbursts~\citep[e.g.,][]{dsw+18}, J1846$-$0258 has not been detected in radio to date. 
Following its magnetar-like X-ray bursts in 2006~\citep{Gavriil2008}, deep observations of PSR~J1846$-$0258 with the Green Bank and Parkes telescopes detected no radio emission \citep{archibald2008ApJ}. This indicated that it is a hybrid object with characteristics of both rotation-powered pulsars and magnetars \citep{Gavriil2008}. PSR~J1846$-$0258 experienced another magnetar-like outburst in August 2020.  After the burst, there were several sharp spikes in pulsed X-ray flux, which returned to static levels in November 2020. The X-ray pulse profile changed during the burst and returned to its pre-burst shape in 2021~\citep{hu2023ApJ, sathyaprakash2024}.

Several searches for radio emissions from magnetars have been conducted previously. \citet{crawford2007ApJ} utilized the Murriyang, CSIRO's Parkes 64-meter telescope for deep observations of four magnetars: 1E 1048.1$-$5937, AX J1845$-$0258, 1E 1841$-$045, and 1RXS J170849.0$-$400910 at a frequency of 1.4 GHz. They reported an upper limit of 0.02 mJy for integrated pulse emission and 1 Jy or better for single pulse flux. In another study, \citet{lazarus2012ApJ} conducted radio observations of five magnetars and two magnetar candidates at 1950 MHz using the Green Bank Telescope between 2006 and 2007. No radio emissions were detected for any of the targets, leading to luminosity upper limits of $L_{1950} \leq$ 1.60 mJy kpc$^2$ for periodic emission and $L_{1950, single} \leq$ 7.6 Jy kpc$^2$ for single pulse emission. \citet{tong2013RAA} observed the magnetar Swift J1834.9$-$0846 with the Nanshan 25-meter Radio Telescope but did not detect any pulsed radio emissions. 
Recently, \citet{lu2024ApJ} used the Five-hundred-meter Aperture Spherical radio Telescope (FAST) to observe three magnetars SGR 0418+5729, 1E 2259+586, 4U 0142+61 and the central compact object (CCO) PSR J1852+0040 at 1.25 GHz. No radio emission was detected through direct either folding or blind search and they provided the strictest upper limits on radio flux for these magnetars and the CCO, ranging from $\lesssim$ 2 to 4 $\mu$Jy.

Here we report a sensitive search for radio emission, both periodic and bursting, from four magnetars SGR 0501+4516, Swift 1834.9-0846, 1E 1841-045, SGR 1900+14 and a magnetar-like pulsar PSR J1846-0258 using the FAST\footnote{http://fast.bao.ac.cn} \citep{nan2011IJMPD, li2018IMMag}. The goal is to detect weak radio emissions from these objects or put tight constraints on the flux of their radio emission and search for FRB-like bursts from radio-quiet magnetars. 
In section\,\ref{sec2}, the observations and data reduction will be presented. 
In section\,\ref{sec3} is the results of the radio pulse searches.
The relevant conclusions and discussions will be given in section\,\ref{sec4}.

\section{OBSERVATIONS} \label{sec2} 
 
Observations of four magnetars SGR 0501+4516, Swift 1834.9$-$0846, 1E 1841$-$045, SGR 1900+14 and a magnetar-like pulsar PSR J1846$-$0258 were conducted using the FAST in 2020. The observations were carried out with the central beam of the 19-beam receiver~\citep{jiang2020RAA}, and exposure time for each source was 2160\,\rm{s} (includes 40\,s of noise injection, an additional 20\,s of overhead, and 2100\,s of observation time for the target source). The FAST ROACH backend was used in its pulsar-search Tracking mode, with 4096 channels across 500\,MHz of bandwidth and 49.152\,$\mu$s sampling rate. The total intensity was recorded with 8-bit sampling, along with full Stokes polarization parameters acquired. The noise diode calibration was used at the beginning of the five observations.  The detailed observation information of each source is listed in Table \ref{table1}.

\begin{table*}
\centering
\caption{Radio search parameters and results}
{\begin{tabular}{lccccc}
\hline
\hline
Parameter  &  SGR 0501+4516  	&  Swift 1834.9$-$0846 	& 1E 1841$-$045  &  SGR 1900+14  &  PSR J1846$-$0258  	        	\\
\hline	
P (s)     &   5.7620695(1)    &   2.4823018(1)    &   11.788978(1)    &    5.19987(7)   &    0.3291956635(5)       \\
$\dot{P}$ (10$^{-11}$ s/s)     &   0.594(2)    &   0.796(12)    &   4.092(15)    &    9.2(4)   &    0.14249(11)       \\
RA                  &   05:01:06.76     &   18:34:52.11     &   18:41:19.34     &    19:07:14.33    &     18:46:24.94       \\
DEC                 &   +45:16:33.92    &   -08:45:56.02    &   -04:56:11.16    &    +09:19:20.1   &     -02:58:30.1      \\
l; b (deg)          &  161.55; +1.95    &   23.25; -0.34    &   27.39; -0.01    &   43.02; +0.77    &   29.71; -0.24    \\
Observation MJD     &   59085.87    &   59178.27    &   59206.23    &   59102.43    &   	59095.47        \\
Length (s)          &   2160    &   2160    &   2160    &   2160    &   2160        \\
Zenith Angle (deg)  &   39.13   &   35.60   &   30.56   &   36.14   &   36.67        \\
G (K Jy$^{-1}$)     &   11.6(3)   &   12.9(3)   &   14.7(2)   &   12.7(3)   &   12.5(3)        \\
T$_{\rm rec}$ (K)       &   30.2(1)   &   29.9(9)   &   28.8(9)   &   29.9(9)     &   30(1)        \\
T$_{\rm 1250,sky}$ (K)  &   2.8      &   18.5     &   16.9     &   13.3       &   18.5         \\
\hline
Distance (kpc)  &   $\sim$ 2    &   4.2(3)    &   8.5$^{+1.3} _{-1.0}$    &   12.5(1.7)    &     6.0$^{+1.5} _{-0.9}$      \\
DM  (YWM16) (pc cm$^{-3}$) &   $\sim$108.99   &  352.03  &  1953.52  & 551.36  & 814.58  \\
Searched DM range (pc cm$^{-3}$) & 0--500  & 0--1100  &  0--6000  & 0--1700  &  0--2500   \\
S$_{\rm 1250, red}$ ($\mu$Jy) & $\lesssim$ 16.9(7)   & $\lesssim$ 8.2(3)  & $\lesssim$ 13.9(3)  &   $\lesssim$ 12.2(4)  &   $\lesssim$ 8.2(3)         \\
S$_{\rm 1250, single}$ (mJy)&  $\lesssim$2.9--89.0  &  $\lesssim$3.8--117.4  &    $\lesssim$3.1--97.3  &  $\lesssim$3.4--106.6   &   $\lesssim$3.9--121.4   \\
L$_{\rm 1250, red}$ ($\mu$Jy kpc$^2$) &  $\lesssim$ 67.6(28)  &  $\lesssim$ 145(21)  &  $\lesssim$ 1000(310)  &   $\lesssim$1900(500)  &  $\lesssim$ 300(150)   \\
L$_{\rm 1250, single}$ (mJy kpc$^2$)& $\lesssim$ 11.6--356.0   & $\lesssim$ 67--2070  &   $\lesssim$ 220--7000  &  $\lesssim$ 530--17000   &   $\lesssim$ 140--4400    \\
\hline
\end{tabular}}
\label{table1}
\end{table*}

\section{Data Analysis and Results}\label{sec3}

\begin{figure} 
   \centering
   \includegraphics[width=6cm, angle=-90]{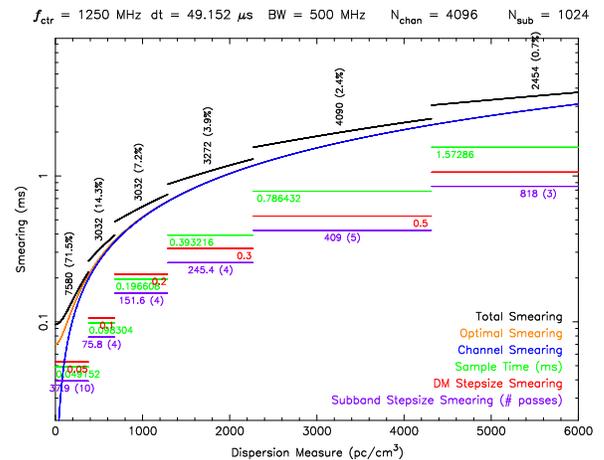}
   \caption{ The de-dispersion plan generated via \texttt{ddplan.py}. }
   \label{ddplan}
\end{figure}

Each observation underwent thorough analysis for periodic radio pulsations and single pulses utilizing the \texttt{PRESTO} software package, as detailed by \citet{ransom2002AJ} and \citet{ransom2011ascl}. The initial stage of both analyses involved the mitigation of narrowband and transient radio frequency interference (RFI) through the \texttt{RFIFIND} command integrated within \texttt{PRESTO}. 
Prior to the search process, data were de-dispersed and then averaged in frequency, encompassing a wide range of Dispersion Measures (DM). The DM values for SGR 0501+4516, Swift 1834.9$-$0846, 1E 1841$-$045, SGR 1900+14, and PSR J1846$-$0258, as predicted by the YMW16 Galactic electron-density model \citep{yao2017ApJ}, together with searched DM ranges were listed in Table~\ref{table1}.
To optimize the de-dispersion accuracy while maintaining computational efficiency, a fine DM spacing was employed for low DM values and a slightly coarser step for higher DM values (DM step sizes in between 0.05--1 pc cm$^{-3}$). This was determined using the  \texttt{DDplan.py} command~\citep{ransom2011ascl}, as illustrated in Figure \ref{ddplan}.

\subsection{Periodic Search}\label{sec3.1}

Given that previous X-ray observations have precisely measured the spin parameters of the four magnetars and the magnetar-like pulsar \citep{camero2014MNRAS, kargaltsev2012ApJ, dib2014ApJ, mereghetti2006ApJ, livingstone2011ApJ}, we extrapolated the period of each source to the current epoch (see Table \ref{table1}) using these X-ray timing results. We then searched for periodic pulsations using the spin periods derived at our observing epochs for the four magnetars and the magnetar-like pulsar, within the same DM range and steps as previously described. This was accomplished by linearly extrapolating the spin period and employing \texttt{prepfold}. The extrapolated spin periods at the start of our radio observations were 5.759819 s for SGR 0501+4516 \citep{camero2014MNRAS}, 2.479967 s for Swift J1834.9$-$0846 \citep{kargaltsev2012ApJ}, 11.77364 s for 1E 1841$-$045 \citep{dib2014ApJ}, and 5.1579 s for SGR 1900+14 \citep{mereghetti2006ApJ}. For PSR J1846$-$0258, the spin period at the start of our radio observations was 0.3291956635 s \citep{hu2023ApJ}. No periodic radio pulsations were found with a significance of S/N = 7.

In addition, we also carried out a blind search of periodic signals in the Fourier domain using \texttt{accelsearch} within the \texttt{PRESTO} software package~\citet{ransom2002AJ}. 
We not only searched for signals from isolated pulsars (with $\texttt{zmax}=0$) but also potential unknown binary systems within our pointing (with $\texttt{zmax}=200$). 
We searched not only for the radio counterparts of these magnetars, but also the presence of any new radio signals at their locations.
Periodic candidates surpassing an S/N threshold of seven were folded with the \texttt{prepfold} command. While no new pulsar candidates were identified, the known PSR~J1907+0918 \citep{Lorimer2000ApJ} was detected in the data obtained from SGR 1900+14.

\subsection{Single Pulse Search}\label{sec3.2}

To search for dispersed single pulses and bursts, we analyzed the dedispersed time series \citep{cordes2003ApJ} using \texttt{single\_pulse\_search.py} from the \texttt{PRESTO} software suite. We used the same DM range and steps as mentioned previously. The native time resolution of 0.05 ms was maintained, and each dispersed time series was searched for candidate single pulses with signal-to-noise ratios (S/N) greater than 7, a threshold set to minimize confusion with RFI background.
To ensure sensitivity to pulses wider than 0.05 ms, we applied matched filtering with boxcar functions of varying widths (from 1 to 30 samples). Additionally, to improve sensitivity to longer-duration single pulses and bursts, the dispersed time series were downsampled by combining original samples into contiguous blocks of 2, 4, 8, 16 and 32 samples. Each set of downsampled time series was then searched using the same boxcar filters, providing sensitivity to signals with durations up to 48 ms (equivalent to the maximum boxcar length of 30 samples multiplied by the maximum downsampled sample time of 1.6 ms).
For each pulsar, we grouped all candidates obtained from the search within a 40-second time interval and a DM interval of 100 pc cm\(^{-3}\) to produce two-dimensional (2D) DM-time plots. These plots were visually inspected to identify significant pulses. Significant candidates were expected to cluster in a high S/N range (greater than 7) near a specific DM value. 
Two single pulses for RRAT J1846$-$0257 \citep{McLaughlin2006Natur} were detected in the pointing towards PSR~J1846$-$0258. We then examined the dynamic spectra of all significant candidates with peak S/N greater than 7. After cross-checking the 2D plots covering the relevant DMs and the dynamic spectra of significant single pulse candidates, we conclude that no astrophysical single pulses were found at an S/N threshold of 7.

\subsection{Upper limit of the flux density}\label{sec3.3}

We can derive upper limits on the flux density for the four magnetars and a magnetar-like pulsar based on the non-detection of periodic radio pulsation and single pulses. Assuming that the data contains only temporally uncorrelated (white) noise, the minimal detectable radio flux density \( S_{\text{min}} \) can be estimated using the radiometer equation \citep{Dewey1985ApJ}:
\begin{equation}
  S_{\rm min,white} = (S/N)_{\rm min} \frac{ \beta (T_{\rm rec}+T_{\rm sky})}{G \sqrt{n_{\rm p} t_{\rm obs} \Delta f}} \sqrt{\frac{W}{P-W}}
  \label{eq1}
\end{equation}
where \((S/N)_{\text{min}}\) is the signal-to-noise threshold required to detect the pulsations, \(\beta\) is a dimensionless factor accounting for quantization losses, \(T_{\text{rec}}\) and \(T_{\text{sky}}\) are the receiver and sky noise temperatures (K), \(G\) is the telescope gain (K Jy\(^{-1}\)), \(n_{\rm p}\) is the number of polarizations summed, \(t_{\text{obs}}\) is the observation duration (s), \(\Delta f\) is the observing bandwidth (MHz), \(W\) is the equivalent width of the pulse profile, and \(P\) is the pulsar period. According to \citet{jiang2020RAA}, \(G\) and \(T_{\text{rec}}\) are functions of the observation zenith angle (\(\theta_{\text{ZA}}\)). \(T_{\text{sky}}\) at 1250\,MHz was estimated based on \citet{Haslam1982A&AS} assuming a spectral index of $-$2.6, as shown in Table \ref{table1}.
In our observation with FAST: \(\Delta f = 500\) MHz, \(n_{\rm p} = 2\), \(t_{\text{obs}} = 2100\) s, and \(\beta = 1\). A pulsed duty cycle of 10\% and a S/N threshold of 7 were assumed for each case. Using Equation \ref{eq1}, we estimate upper limits on the flux density at 1250 MHz for the four magnetars SGR 0501+4516, Swift J1834.9$-$0846, 1E 1841$-$045, SGR 1900+14 and the magnetar-like pulsar PSR J1846$-$0258 to be 4.6, 6.0, 5.0, 5.5, and 6.2 $\mu$Jy, respectively. 

For long-period pulsars, the detectability of pulsations can be significantly affected by temporally correlated (red) noise in the data as discussed by \citet{Lazarus2015ApJ}. 
To account for the red noise, we folded each observation using the expected DM of the source (Table~\ref{table1}) at two different periods: the source's period and a 10\,ms period. Identical RFI channels were masked for both folded files, which were then averaged in time and frequency to generate pulse profiles. Profiles folded at 10\,ms showed only white noise, whereas profiles folded at the magnetar and pulsar periods displayed significant red noise. We calculated the RMS noise of each profile ($\sigma_{\rm 10ms}$ and $\sigma_{\rm p0}$) and estimated the upper limit on flux density as $S_{\rm min,red} = S_{\rm min,white}\times (\sigma_{\rm p0}/\sigma_{\rm 10ms})$.
After accounting for the effects of red noise, revised upper limits on the flux density are: 16.9 $\mu$Jy for SGR 0501+4516, 8.2 $\mu$Jy for Swift J1834.9-0846, 13.9 $\mu$Jy for 1E 1841-045, 12.2 $\mu$Jy for SGR 1900+14, and 8.2 $\mu$Jy for PSR J1846$-$0258.

The upper limit of a single pulse can be calculated using 
\begin{equation}
  S_{\rm SP} = (S/N)_{\rm min} \frac{ \beta (T_{\rm rec}+T_{\rm sky})}{G} \sqrt{\frac{1}{n_{\rm p}  \Delta f W}}
  \label{eq2}
\end{equation}
By applying the modified radiometer equation (Equation \ref{eq2}) and the above S/N threshold, we determined the detection upper limits for single-pulse flux density and luminosity across the range of pulse timescales 0.05–48 ms. These limits are presented in Table \ref{table1}, with values between 2.9 and 135.9 $\mu$Jy.

\section{Discussion and Conclusions}\label{sec4}

\begin{figure} 
   \centering
   \includegraphics[width=9cm]{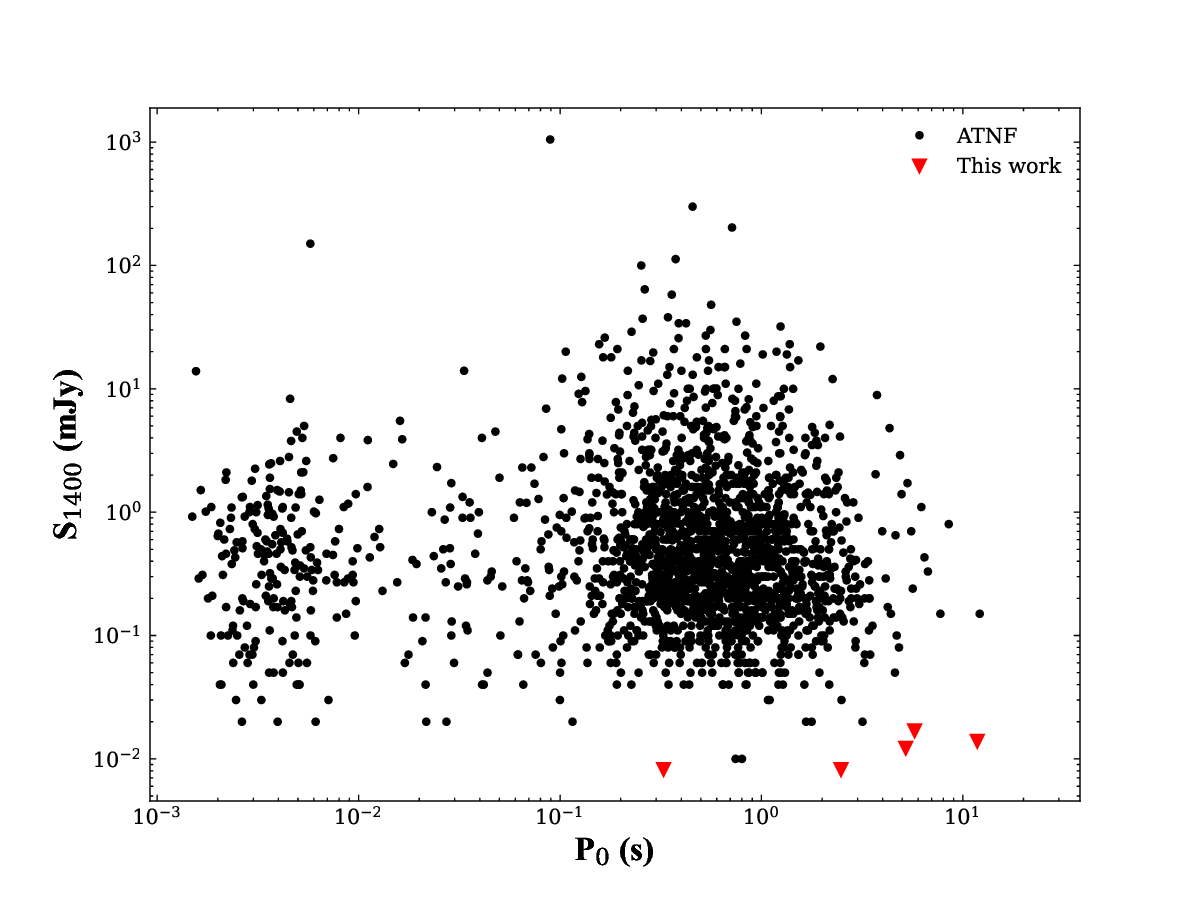}
   \caption{The flux density of radio pulsars at 1.4 GHz as a function of the spin periods. Black points represent pulsars from the Australia Telescope National Facility (ATNF) Pulsar Catalogue \citep{Manchester2005AJ}. Red lower triangles denote the upper limits of periodic pulsation from the four magnetars and a magnetar-like pulsar in this work. }
   \label{P0_S1400} 
\end{figure}

We conducted deep radio searches for pulsations in four magnetars SGR 0501+4516, Swift J1834.9$-$0846, 1E 1841$-$045, SGR 1900+14 and a magnetar-like pulsar PSR J1846$-$0258. No periodic radio emission or single pulses were detected during our observation epochs. Our results indicate that any periodic or bursting radio emission from these sources is either turned off, extremely weak, or very sporadic during the observation periods. After taking into account the red noise, we derived upper limits on the radio flux density to be 16.9 $\mu$Jy for SGR 0501+4516, 8.2 $\mu$Jy for Swift J1834.9$-$0846, 13.9 $\mu$Jy for 1E 1841$-$045, 12.2 $\mu$Jy for SGR 1900+14, and 8.2 $\mu$Jy for PSR J1846$-$0258.  These upper limits on flux density are at least several times lower than measured radio flux densities (at $\sim1.4$\,GHz) of known long-period pulsars (e.g., $P_0>1$\,s, see Figure \ref{P0_S1400}). The range of single-pulse 1250 MHz flux limits for the timescales searched (0.05--48 ms) is presented in Table \ref{table1}. The 1250 MHz luminosity limit on pulsed emission and single-pulse flux limits are also detailed in Table \ref{table1}.

The magnetars that have shown pulsed radio emission thus far have been observed to have rotational energy loss rates that exceed their X-ray luminosity during quiescence, with high-energy bursts triggering the radio emission~\citep{rea2012ApJ}. The quiescent X-ray luminosities of both Swift J1818.0$-$1607 and SGR 1935+2154 are less than their rotational energy loss rate~\citep{Hu2020ApJ, Borghese2020ApJ}. Recently, both Swift J1818.0$-$1607 and SGR 1935+2154 have been detected in radio after their X-ray bursts~\citep{Lower2020ApJ, andersen2020Natur, Zhu2023SciA}, which supports the fundamental plane of magnetar radio emission from \citet{rea2012ApJ}. During our observations, no bursts were detected at high energies for the four magnetars SGR 0501+4516, Swift J1834.9$-$0846, 1E 1841$-$045, and SGR 1900+14. For the magnetar-like pulsar PSR J1846$-$0258, which experienced a burst announced by Swift on August 1, 2020, the pulse flux increased sharply several times post-eruption, stabilizing by November 2020 \citep{hu2023ApJ}. Our observations took place on September 03, 2020, nearly one month after the X-ray burst, and at that time, the pulsed X-ray emission had not yet returned to resting levels. For PSR~J1846$-$0258, assuming a distance of 6.0\,kpc, the quiescent X-ray luminosity is approximately \(L_X \sim 1.9 \times 10^{34}\) erg s\(^{-1}\). The spin down luminosity of this pulsar is around \(8.05 \times 10^{36}\) erg s\(^{-1}\), resulting in an \(L_X/L_{\text{rot}}\) ratio of about \(2.36 \times 10^{-3}\), which is significantly less than unity. This similarity to other radio-emitting magnetars suggests that PSR~J1846$-$0258 is likely to exhibit radio emission if the scenario proposed by \citet{rea2012ApJ} holds. 
Considering that our observation only lasted approximately 30 minutes, it might be insufficient to conclusively rule out the presence of radio emissions. More extensive radio monitoring following the X-ray burst is required.

The absence of radio emission in magnetars can be attributed to several factors, including suppression by strong magnetic fields, the directionality of radio beams, distinct emission mechanisms, environmental influences, and the limitations of current observational techniques. Magnetars possess extremely strong magnetic fields, typically ranging from $10^{14}-10^{15}$ G, which can suppress the mechanisms responsible for generating radio emissions. These strong magnetic fields can also affect the propagation of electromagnetic waves, potentially leading to the absorption or scattering of radio waves \citep{thompson1995MNRAS}. Additionally, any radio emission from magnetars might be highly directional. If the radio beam is not aligned towards the Earth, it would remain undetectable by terrestrial observers \citep{beloborodov2007ApJ}. The emission mechanisms in magnetars differ from those in regular pulsars. While pulsars typically emit radio waves due to particle acceleration along magnetic field lines, magnetars primarily emit X-rays and gamma rays due to magnetic reconnection and charge separation events \citep{rea2011ASSP}. Furthermore, the environment surrounding a magnetar can impact the propagation of radio waves. High-density plasma around the magnetar could absorb or scatter radio emissions, making them challenging to detect \citep{camilo2006Natur}. Current radio telescopes may not have the sensitivity needed to detect weak radio signals from magnetars. Additionally, radio emissions from magnetars might be transient or intermittent \citep{kouveliotou2003SciAm, burgay2009ATel}, necessitating prolonged monitoring or high-time-resolution observations to capture them.

\section*{Acknowledgments}

This work is supported by the National Natural Science Foundation of China (No. 12288102), the Major Science and Technology Program of Xinjiang Uygur Autonomous Region (No. 2022A03013-3), the National Natural Science Foundation of China (No. 12041304, 12273008, 12041303, 12041304, 12403046), the National SKA Program of China (Nos. 2022SKA0130100, 2022SKA0130104, 2020SKA0120200),  The research is partly supported by the Operation, Maintenance and Upgrading Fund for Astronomical Telescopes and Facility Instruments, budgeted from the Ministry of Finance of China (MOF) and administrated by the Chinese Academy of Sciences (CAS), the Natural Science and Technology Foundation of Guizhou Province (No. [2023]024), the Foundation of Guizhou Provincial Education Department (No. KY (2020) 003), the Academic New Seedling Fund Project of Guizhou Normal University (No. [2022]B18)  and the Scientific Research Project of the Guizhou Provincial Education (Nos. KY[2022]137, KY[2022]132). 
This work made use of the data from FAST (Five-hundred-meter Aperture Spherical radio Telescope) (\url{https://cstr.cn/31116.02.FAST}). FAST is a Chinese national mega-science facility, operated by National Astronomical Observatories, Chinese Academy of Sciences.

\bibliography{references}{}

\bibliographystyle{aasjournal}

\end{document}